\documentclass[aps,prl,twocolumn,groupedaddress]{revtex4}
\usepackage{amssymb}
\usepackage{graphicx}  

\newcommand{\ket}[1]{\mbox{$ | #1 \rangle $}}

\begin{document}

\title{Quantum teleportation over the Swisscom telecommunication network}

\author{Olivier Landry \footnote{email: olivier.landry@physics.unige.ch}}

\author{J.A.W. van Houwelingen}
\author{Alexios Beveratos}
\author{Hugo Zbinden}
\author{Nicolas Gisin}

\affiliation{Group of Applied Physics, University of Geneva, 20 rue de
  l'\'Ecole-de-m\'edecine, 1205 Gen\`eve, Switzerland}

\begin{abstract}
We present a quantum teleportation experiment in the quantum relay
configuration using the installed telecommunication network of
Swisscom. In this experiment, the Bell state measurement occurs well
after the entanglement has been distributed, at a point where the photon
upon which data is teleported is already far away, and the
entangled qubits are photons created from a different crystal and laser
pulse than the teleported qubit. A raw
fidelity of 0.93$\pm$0.04 has been achieved using a heralded single-photon source.
\end{abstract}

\maketitle 

\section{Introduction}

Quantum teleportation, or the ability to transfer information in the
form of qubits between two
locations without a direct quantum channel, has many practical applications
in addition to its fundamental significance. In quantum communication protocols, such as quantum key distribution
\cite{Bennett1992a, Gisin2002}, the maximum distance one can reach is
limited by channel losses and detector noise. When the signal to noise
ratio gets lower than a certain limit which depends on the choice of
protocol and experimental details, no secure information can be
retrieved. While the losses in optical fibers cannot be decreased with current
technologies, a suitably designed communication channel making
use of quantum relays\cite{Jacobs2002, Collins2005} or quantum
repeaters\cite{Brugiel1998} can reduce the noise and therefore
increase the distance limit. Quantum relays uses quantum
teleportation and entanglement swapping to perform a kind of quantum
non-demolition measurement at some points within the channel, in
effect measuring the presence of a photon without measuring the qubit
it carries. Detectors can then be opened only when a photon is certain
to arrive, reducing dark counts, the main source of noise. Quantum repeaters perform the same task but also use
quantum memories and entanglement purification. 

In previous experiments,
teleportation was demonstrated inside a
laboratory\cite{Bouwmeester1997,Popescu1998,Marcikic2003} or in the field but without prior
entanglement distribution\cite{Ursin2004}. The Bell-State
Measurement always took place before the third photon was distributed
to Bob. On the other hand, in all these experiments
the same laser pulse was used to create both the entangled pair
and the photon to be teleported. These two points limit the
feasibility of a practical quantum relay and open conceptual loopholes.
Here we present an experiment where teleportation occurs long after
entanglement distribution and the photons involved originate from two
crystals excited by different pulses from the same laser.

\section{Protocol}

\begin{figure}[htbp] 
\centering
\includegraphics[width=0.4\textwidth]{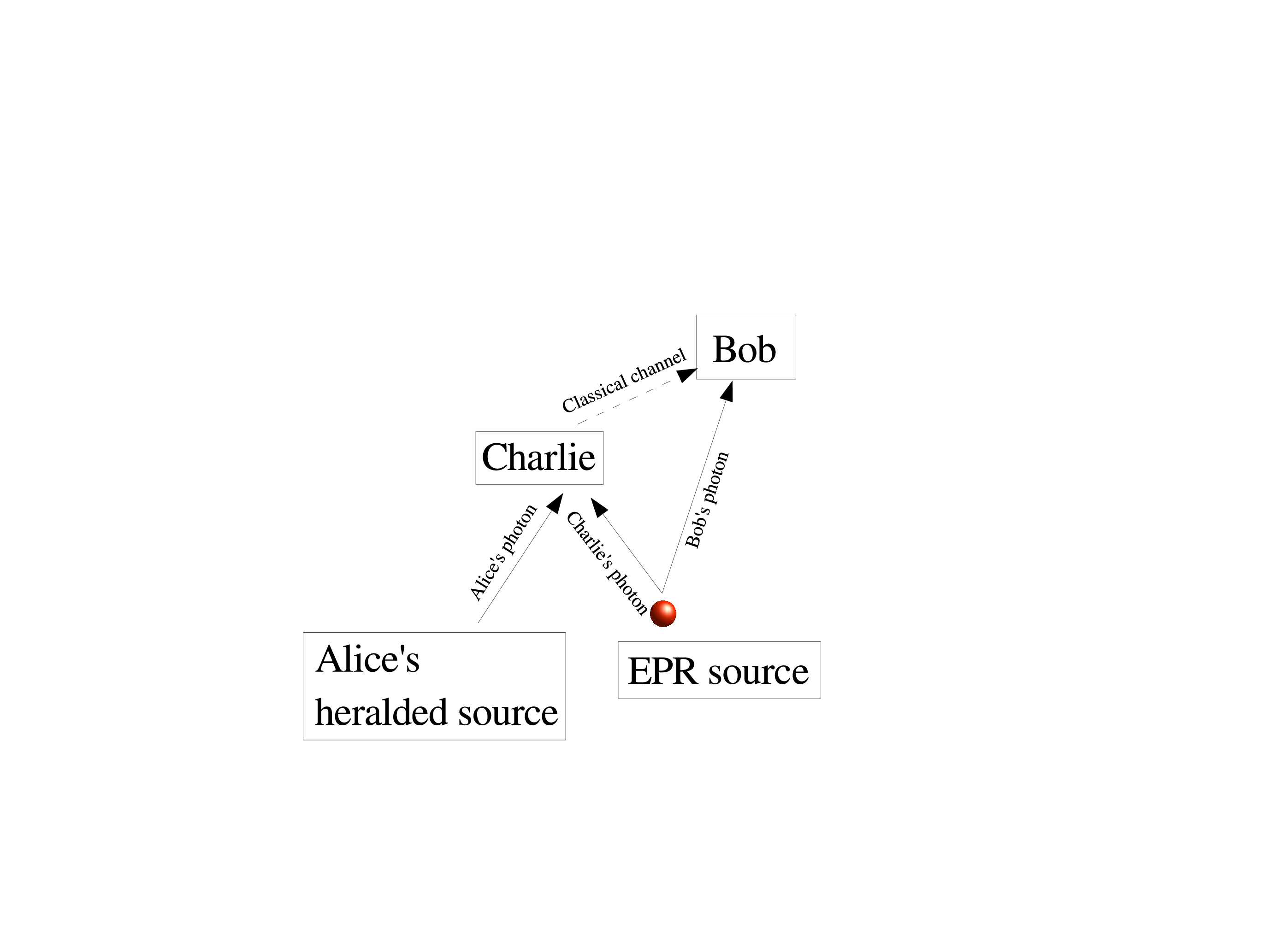}
\caption{\label{concept}The teleportation protocol. Quantum channels
  are in plain lines, classical channels are in dashed lines. The state that Bob
  measures is the same that Alice sent to Charlie up to a unitary transformation. EPR source is a source of entangled states, or
  Einstein-Podolsky-Rosen states \protect\cite{Einstein1935a}. 
}
\end{figure}

The quantum teleportation protocol \cite{Bennett1993} (schematically
described in fig \ref{concept}) requires that Bob (the
receiver) and Charlie (a third party) share an entangled state, which
in this case is a $\ket{\phi^{+}}$ state.
Alice (the sender) needs to
send a qubit over to Bob, but does not possess a direct quantum channel. She
sends it to Charlie who performs a Bell State measurement\cite{Houwelingen2006}
using a beamsplitter and classically
announces the result to Bob. 

\subsection{Time-bin qubits}

Time-bin qubits \cite{Brendel1999} have proven to
be very robust against decoherence in optical fibers\cite{Thew2002} and several
long distance experiments have been demonstrated\cite{Marcikic2003,
  Riedmatten2004, Riedmatten2005}. In this experiment time-bin qubits
lying on the equator of the Bloch sphere are created using an
unbalanced interferometer. For a review of time-bin qubits see Tittel
and Weihs\cite{tittel2001}.

\section{Setup}

\begin{figure}[htbp] 
\centering
\includegraphics[angle=0, width=0.4\textwidth]{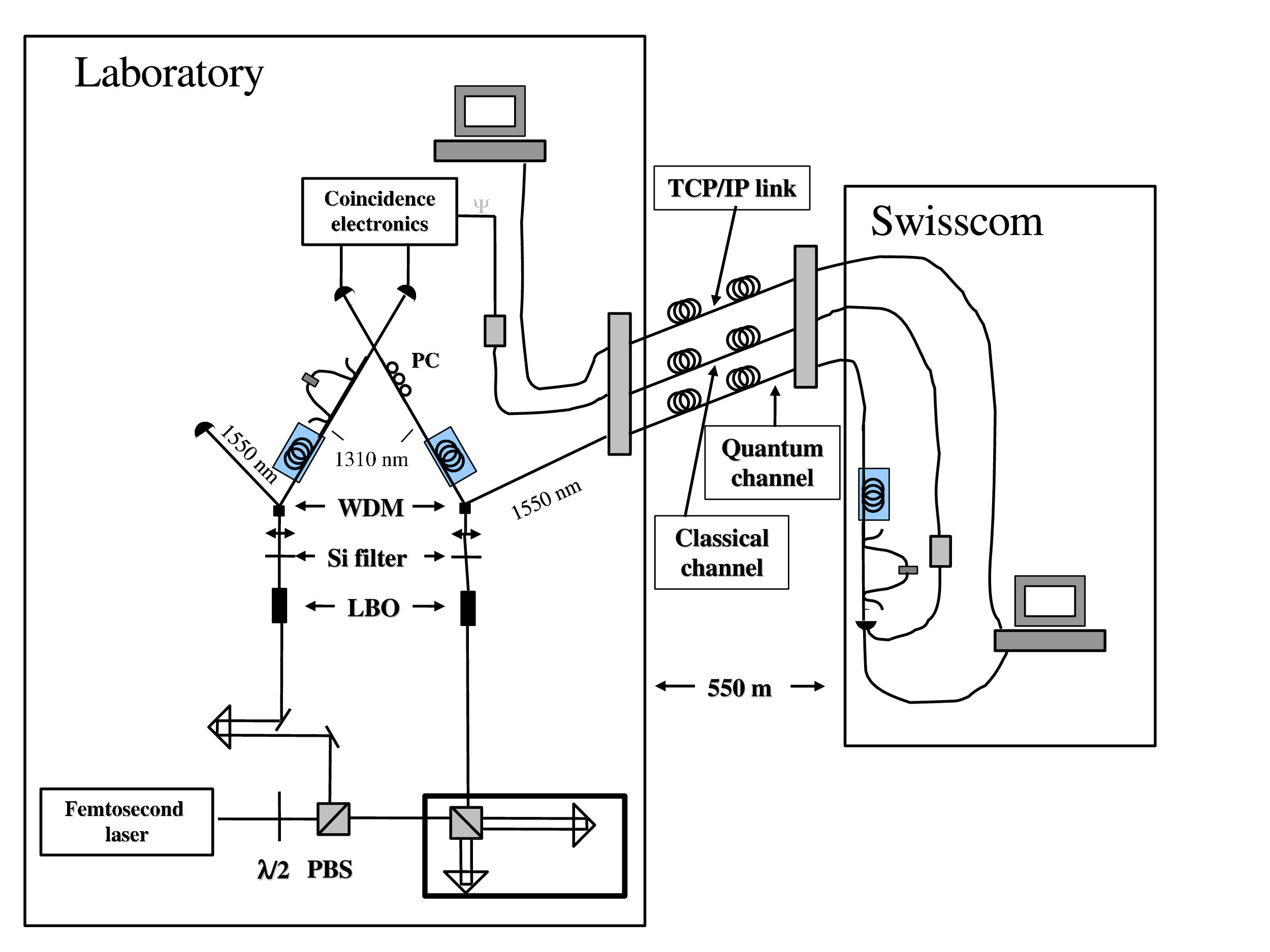}%
\caption{\label{circuit}The optical circuit. PC: Polarization
  controller. QM: rudimentary Quantum Memory (fiber spools). }
\end{figure}

The experimental setup is shown in fig. \ref{circuit}. A mode-locked
Ti:Sapphire laser (Mira Coherent, pumped using a Verdi laser)
creates 185\,fs pulses with a spectral width of 4\,nm at a central
wavelength of 711\,nm, a mean power of 400mW and a repetition rate of 75\,MHz. This beam is split
in two parts using a variable coupler ($\lambda/2$ and a polarization
beamsplitter). 

The
transmitted light is sent through an unbalanced Michelson
interferometer stabilized using a frequency-stabilized HeNe laser
(Spectra Physics 117A) and then on a Lithium Borite (LBO) non-linear crystal (NLC) cut for
type-I phase-matching, which creates a time-bin entangled photon pair in the $\ket{\phi^{+}}$ state
by spontaneous parametric downconversion. The created photons have wavelengths of 1310 and 1555\,nm and are easily separated using a
wavelength division multiplexer (WDM). A Si filter is used to remove
the remaining 711\,nm light.

\subsection{Charlie's photon}

The 1310\,nm photon is sent in a 179.72\,m spool of fiber. This spool
serves as a rudimentary quantum memory (QM). While Charlie's part of the
entangled qubit pair is waiting in this spool, Bob's part leaves the
laboratory. 

\subsection{Alice's photon}

Alice prepares her photon using the light reflected from the
variable coupler. A pair of photons is created in the same type
of crystal as above, then separated. The 1555\,nm photon can
either be discarded or detected by an InGaAs APD in order to herald the photon to
be teleported\cite{Fasel2004, Pittman2004}. If it is not detected,
teleportation still occurs without other changes in the setup.

The
1310\,nm photon is stored in a 177\,m spool of fiber. The 2.72\,m
difference with Charlie's QM corresponds exactly to the spacing
between two subsequent pulses of the laser. This means that
Alice's photon upon which the qubit to be teleported will be
encoded is produced from a different pulse of the laser than
Charlie's and Bob's photons. 
This is a conceptually important step towards
completely independant laser sources\cite{Yang2006, Kaltenbaek2006}.

In order to encode a qubit on her photon, Alice sends it after the
spool to an unbalanced fiber interferometer independantly and actively stabilized
\cite{Marcikic2003} by a frequency stabilized laser at a
wavelength of 1552\,nm (Dicos OFS-2123). Only then is Alice's qubit created.
Note that at this point, Bob's photon is already 177\,m away from
the laboratory.

Once Alice's photon has been encoded, Charlie performs a Bell-State
measurement (BSM)
jointly with his photon and the photon Alice prepared.

\subsection{Alignment and Stabilization}

Alice's and Charlie's photons need to arrive at the
beamsplitter within their coherence time and be indistinguishable for
the Bell state measurement to
be successful. Charlie's photon passes through a polarization
controller to make both polarizations equal at the beamsplitter.
Chromatic dispersion is
negligible at the 1310\,nm wavelength.
Charlie filters both photons
down to 5\,nm of bandwidth, which corresponds to a coherence time of
$\tau_{c}$=500\,fs or a coherence length of $L_{c}$=150\,$\mu$m,
approximately three
times more than the excitation pulse. 
The easiest way to control the distance traveled by the photons with
this precision is add a variable delay consisting of a retroflector mounted on a micrometer step
motor placed right after the variable coupler which can
move with a precision of 200\,nm.

\begin{figure}[htbp] \centering
\includegraphics[angle=0, width=0.4\textwidth]{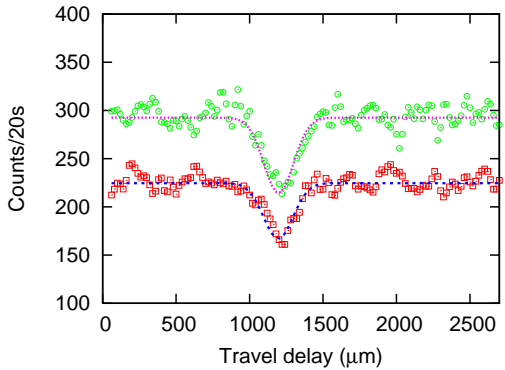}%
\caption{\label{mandel}The circles corresponds to the number of
long path coincidences in the coupler, the squares to short path
coincidences. Both show a Mandel dip (with visibilities $V_{11}=0.266$ and
$V_{00}=0.255$ respectively), which demonstrates
indistinguishability. Both Mandel dips also are at the same location,
demonstrating good alignment of the interferometers. }
\end{figure}

A Mandel Dip \cite{Hong1987} experiment (fig. \ref{mandel}) is performed in
order to measure the degree of temporal indistinguishability between the two
incoming photons. The registered raw
visibilities are $V_{00}=0.255$ and $V_{11}=0.266$
for the short and long paths respectively. 
Corrected visibilities are
near to the theoretical maximum
net visibility of $\frac{1}{3}$\cite{Riedmatten2003}. The
width of the Mandel Dip, which corresponds to the coherence length of
the photons, is 144\,$\mu$m as expected for filters of 5
\,nm. 
Since the fiber spool at Alice's side is longer
by one pulse period, the two photons that
experience photon bunching have not been created by the same
excitation pulse but from subsequent ones.

Unfortunately, the alignment of the different paths is not inherently stable. Temperature
fluctuations in the laboratory will affect the length of the
fibers and the repetition rate of the laser. In order to avoid
length fluctuations of the fiber spools, they have been placed in
a common insulated box so that any fluctuation will apply equally
to both. The fluctuation in the length difference of the nominally
equal 177\,m spools of fiber has been measured in a Mach-Zender
setup to be less than 10\,$\mu$m. Longer fibers similarly insulated
show larger fluctuations: spools of 800\,m showed fluctuations of
up to 60\,$\mu$m over less than an hour, which would have destroyed 
complete indistinguishability on the long term.

There remains two effects that have to be compensated. First, the
thermal expansion of the additional 2.72\,m of fiber on Charlie's side
is not compensated by an equal length on Alice's side as for the other
fibers. This 2.72\,m will undergo a thermal expansion of
$\sim$100\,$\mu$m/K. Secondly, the laser repetition rate fluctuates in
a seemingly random fashion by up to
400\,Hz/h in the worst cases, a fluctuation which would cause a change
of 15\,$\mu$m in the additional length needed to skip exactly one
period of the laser. These phenomena combined mean that Alice's
photon and Charlie's photon will not stay indistinguishable for more
than a few hours, not enough to perform a teleportation experiment.

\begin{figure}[htbp] 
\centering
\includegraphics[angle=0, width=0.4\textwidth]{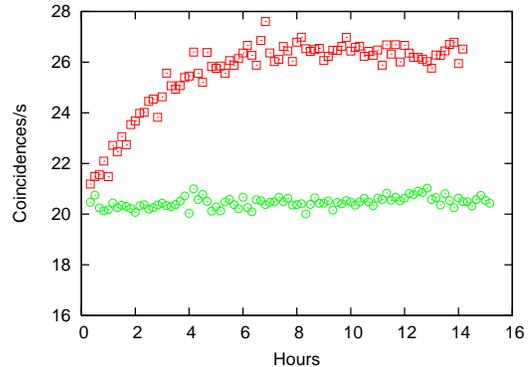}%
\caption{\label{stability}In squares, the number of counts (sum of short and
  long paths, normalized) without active stabilization. After a few hours, the
  coincidence rate has risen to a level equal to that outside the
  Mandel dip. This means that the photons path lengths are different
  by more than their coherence length and no bunching occurs. In circles, a similar experiment with active
  stabilization shows that the indistinguishability condition is now
  stable. landryfig4.eps}
\end{figure}

The squares curve in fig. \ref{stability} demonstrates this instability. A Mandel
dip experiment is performed to find the minimum of the curve as in
fig. \ref{mandel}. The motor
moves to this point at time zero and is not moved afterwards. It can
clearly be seen that the number of coincidences registered increases with
time. After a few hours, a plateau is reached where no bunching occurs
anymore.

Numerous experiments have shown that the departure from the
minimum of the Mandel dip is very well correlated with the repetition rate of the
laser which is registered using an external counter (Agilent
53131A). This is because the only parameter involved is the
temperature of the laboratory; the laser can then be used as a
very sensitive thermometer. It is therefore possible to link the
stepmotor movement with the measured repetition rate using a LabView
program such that the motor moves at an empirically derived rate of
0.07\,$\mu$m/Hz. 
In this way, the step motor position is constantly adjusted for
optimal indistinguishability.
The
results are also shown in fig. \ref{stability} in circles.

Another way of stabilizing the length difference would have been to measure
the number of coincidences when inside the Mandel Dip and move the
motor accordingly in a PID feedback loop. However, due to the low
count rate the integration time would have been longer than the
observed fluctuation time, making such a system inefficient.

\subsection{Bell-State Analyzer and electronics}

Both 1310\,nm photons are sent to a Bell State Analyzer (BSA) consisting
of a beam splitter and two detectors (Charlie). One detector is a
passive Ge Avalanche Photodiode (APD), the other is a InGaAs APD
(IdQuantique Id 200) triggered by the first one.

\begin{figure}[htbp] 
\centering
\includegraphics[angle=0, width=0.4\textwidth]{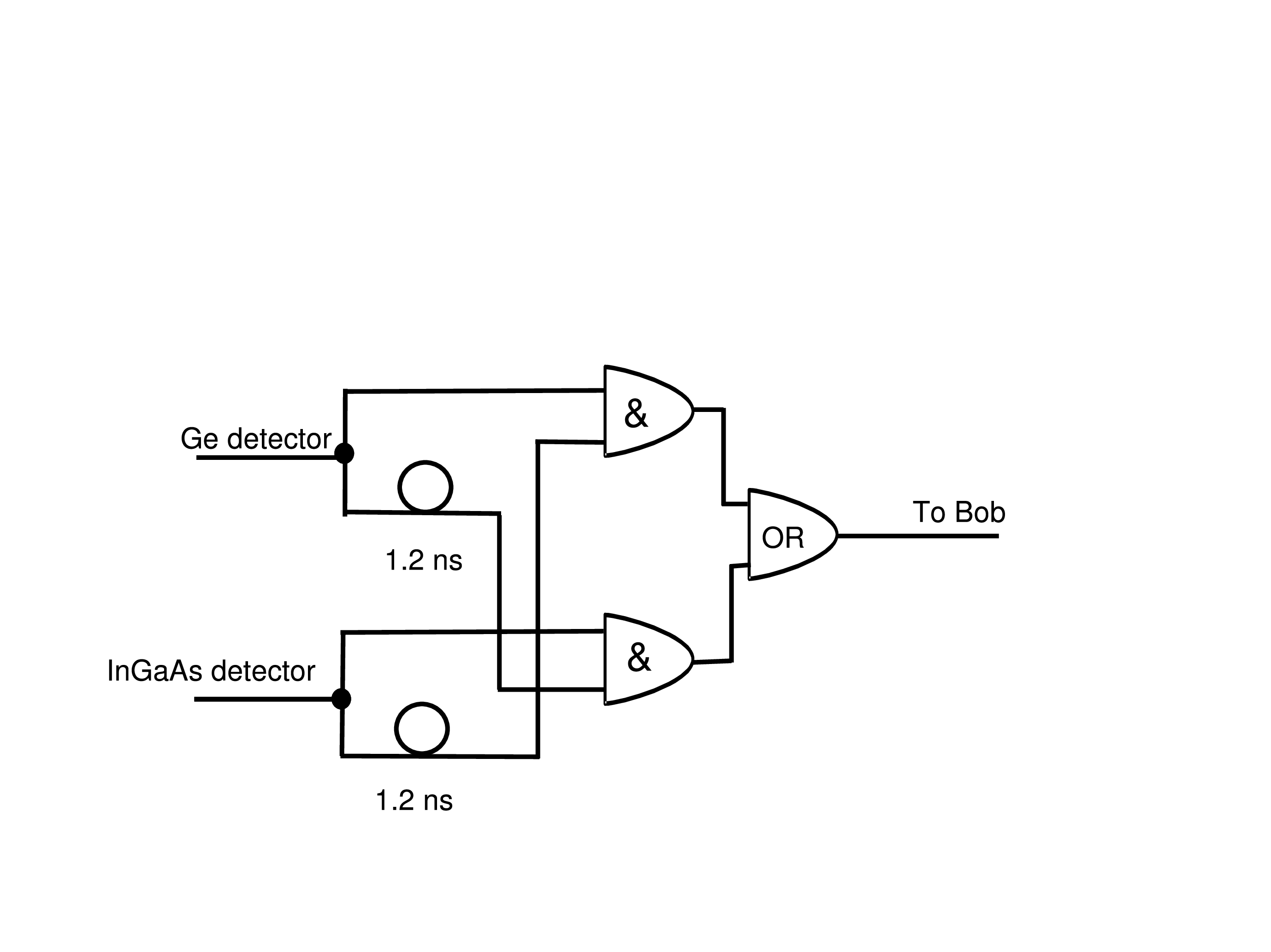}%
\caption{\label{electronics}A schematic of the electronic circuit used
to compare the photons time of arrival in the BSA. Each time a photon
is detected in each detector with a 1.2 ns delay between each photon,
which corresponds to a $\ket{\Psi^{-}}$ detection, a triggering signal
is sent to Bob. }
\end{figure}

An electronic signal is generated  for each photon
detection. These signals time arrivals are electronically compared as
shown in fig. \ref{electronics}. A
$\ket{\Psi^{-}}$ detection occurs when two photons arrive on different detectors
with a time difference of one time-bin, or 1.2\,ns in our setup. To reduce
dark counts, only those
photons which are coincident with a clock signal from the laser are
considered. In total, the electronic circuit is able to make a
decision whether the signal is a $\ket{\Psi^{-}}$ in about
220\,ns. This time corresponds to the physical cable lengths and various
delays that had to be implemented to trigger the active
detectors and synchronize and transform the electric signals. When the conditions for a $\ket{\Psi^{-}}$ are not
fulfilled, no information about the state is available.
 When a $\ket{\Psi^{-}}$ has been
successfully detected, the information is sent to Bob over a
second optical fiber (the classical channel) and by means of an
optical pulse.

\subsection{Bob's photon}

Bob is at a Swisscom substation at a flight
distance of 550\,m from the laboratory, but an optical fiber distance
of 800\,m. Losses in the these fibers are smaller than 2\,dB. To
minimize chromatic dispersion and reduce spurious detections, the photon is filtered down to a 15\,nm
width. 

Upon
receiving the qubit, Bob stores it in a very basic quantum memory
consisting of a fiber spool of 250\,m waiting for the BSA's
information to arrive. Once the confirmation of a successful
$\ket{\Psi^{-}}$ measurement reaches Bob, he opens his InGaAs APD
(IdQuantique Id 200) and detects
the incoming photon after sending it through an analyzing
interferometer. Photons which do not correspond to a successful
$\ket{\Psi^{-}}$ measurement from Charlie are discarded. Arrival times of the photons with respect to the
classical signal are measured by means of a Time to Digital Converter
(TDC). Timing jitter between the classical information and the
qubit is negligible and much smaller than the timing resolution of
the InGaAs detector.

By scanning the analyzing interferometer, Bob measures the
visibility of the interference in order to extract the fidelity of
the teleported state. Bob actively stabilizes his interferometer with a
frequency stabilized laser at a wavelength of 1533\,nm (Dicos OFS-320).

It is important to note that Bob is a completely independant
unit, with its own local interferometer stabilization and
controls. 
A LabView program developped in-house allows an operator to control
Bob, Charlie and Alice from the laboratory using a dedicated TCP/IP
channel that uses the third fiber indicated in
fig. \ref{circuit}. In particular, the detectors have to be closed
when the interferometers are being stabilized; synchronization of the
stabilization and measurement periods are made through this
channel. Crosstalk between this fiber and the quantum channels is
negligible, even at the single-photon level.
 
\subsection{Difference between three-photon and four-photon setup}

The laser light does not need to be separated equally between the two
NLCs and the ratio can be adjusted to minimize noise\cite{Marcikic2003}.
When the 1555\,nm photon from Alice's NLC is not detected, double
pair creation in the EPR source can create a false signal even if
there is no photon created in Alice's NLC since it is possible that one
photon from the double pair will find its way to the Ge APD and the
other one to the InGaAs APD. On the other hand, double pair production
in Alice's NLC will not be recorded if there is no corresponding
photon at Bob's. Therefore in this case it is usual to use less power on the EPR
source than on Alice's source to minimize the number of false
counts. However, when the 4$^{\textrm{th}}$ photon is also
detected, no false signal will be recorded unless there is also a pair
created at Alice's source, therefore we can use equal power on both
sources. The resulting noise reduction allows a greater
signal-to-noise ratio.

\section{Results}

\subsection{Teleportation without heralded single-photon}

\begin{figure}[htbp] 
\centering
\includegraphics[angle=0, width=0.4\textwidth]{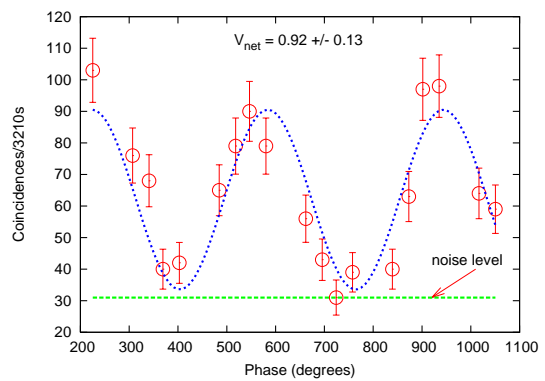}%
\caption{\label{diffpulses} A teleportation experiment: as we scan the
phase of Bob's interferometer, the number of coincidences oscillates,
which shows that the qubit is in the expected superposition state. When
substracting noise, the near-perfect visibility $V_{net}=0.92 \pm 0.13$ shows that decoherence
is minimal. The raw visibility of $V_{raw}=0.46\pm0.06$ is still higher than
the classical limit. }
\end{figure}

A first experiment was performed without detecting the 4$^{\textrm{th}}$ photon
to allow a greater count rate. We first performed a Mandel dip to adjust
the variable delay and linked its position to the repetition rate of
the laser as described before. We
locked the phases of the bulk and Alice's interferometer and slowly
scanned Bob's interferometer phase. Each point was measured for 53\,min
to accumulate statistics. The results are shown in
fig. \ref{diffpulses}.

The power on the entangled photon source was lowered compared to the
power on Alice to
reduce noise. The probability of creating a pair of photon per pulse on
Alice's NLC was $P_{A}=0.19$ and the probability of creating a pair of
entangled photon per pulse was $P_{B}=0.07$. 

The presence of two complete periods shows that the expected cosine is well
reproduced. The visibility of the curve and the fidelity\cite{Riedmatten2004} $F=(1+V)/2$
are good measures of the quality of the teleportation.
The raw visibility
$V_{raw}=0.46\pm0.06 (F_{raw}=0.73\pm0.03)$ is higher than the
classical limit $V=\frac{1}{3} (F=\frac{2}{3})$. The difference with a
perfect visibility comes mainly from known sources of noise such as
double pair production in the NLC and dark counts in the detectors. 
This noise can be measured by running the same experiment but blocking
different parts of the signal.
When these sources of noise are measured
and substracted, we get a net visibility of $V_{net}=0.92 \pm 0.13$
($F_{net}=0.96 \pm 0.06$). We can conclude that decoherence is minimal.

\subsection{Teleportation with heralded single-photon}

\begin{figure}[htbp] 
\centering
\includegraphics[angle=0, width=0.4\textwidth]{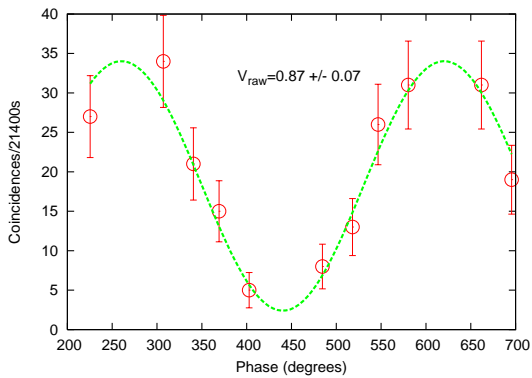}%
\caption{\label{4photon} When taking into account the 4$^{th}$ photon,
noise is greatly reduced and we are able to see a raw visibility of
$V_{raw}=0.87\pm0.07$. }
\end{figure}

Even though the sources of the reduced visibility are known and
understood, for practical applications a high raw visibility is
needed. We performed a second experiment where we detected the
4$^{\textrm{th}}$ photon to transform Alice into a heralded single-photon source.
In
this case, the probability of creating a pair was set to be roughly
equal in both NLCs at $P=0.13$.  
The efficiency of the detector and additional losses induced by the
additional optical components meant that the overall signal was
reduced by a factor 15, and each point necessitated 6\,hrs of data accumulation. However the noise reduction meant
that the raw visibility was much higher at $V_{raw}=0.87\pm0.07
(F_{raw}=0.93\pm0.04)$, which is higher than the cloning limit
$V=\frac{2}{3} (F=\frac{5}{6})$\cite{Grosshans2001, Bruss1998a}. The
results are shown in fig. \ref{4photon}.

In this case, the noise is so low as to be unmeasurable by
conventional means. Therefore, we did not measure the net
visibility. We should point out that in this case, the raw visibility
is as high (within uncertainties) as the net visibility of the
previous experiment. We can conclude that the main sources of noises
can be eliminated by using a heralded single-photon source.

\section{Conclusion}

In summary, we have performed a teleportation in conditions that are
close to field conditions. Bob was a completely independant setup and was
remotely controlled. Alice's qubit was created only after the
entanglement distribution took place. The necessary optical delays were
stabilized using the variations of the repetition rate of the laser,
an easily obtainable information. The qubits were created using
different pulses of this laser. In the future, using truly independant
lasers will enable the construction of a teleportation machine which
will have the ability to receive and transport information from
another independant machine, an ability which would be helpful in quantum
networks for example. Using a
heralded single-photon source we were able to
obtain a raw fidelity of $F=0.93\pm0.04$.

\section*{Acknowledgments}
The authors thank Jean-Daniel Gauthier and Claudio Barreiro for their
support with the electronic components of this project. This work has been supported by the European Commission under the
IST Integrated Project ``Qubit Applications'' (QAP) and by the
Swiss NCCR Project ``Quantum Photonics''. The authors would like to thank
Swisscom for generously giving access to their network and facilities.













\end{document}